# General formula for the efficiency of quantum-mechanical analog of the Carnot engine

**Sumiyoshi Abe**

Department of Physical Engineering, Mie University, Mie 514-8507, Japan

**Abstract:** An analog of the Carnot engine reversibly operating within the framework of pure-state quantum mechanics is discussed. A general formula is derived for the efficiency of such an engine with an arbitrary confining potential. Its expression is given in terms of an energy spectrum and shows how the efficiency depends on a potential as the analog of a working material in thermodynamics, in general. This nonuniversal nature results from the fact that there exists no analog of the second law of thermodynamics in pure-state quantum mechanics where the von Neumann entropy identically vanishes. A special class of spectra, which leads to a common form of the efficiency, is identified.

Thermodynamics of small systems is of contemporary interest, ranging from (bio)molecular and nano scales to even scales of a few particles. From the viewpoint of energetics, one of primary issues is to understand how the works are extracted from such systems. Just a little more than a decade ago, a pure-state quantum-mechanical analog of the Carnot engine reversibly operating at vanishing temperatures has been proposed in Ref. [1]. It is the smallest engine consisting only of a single quantum particle confined in the one-dimensional infinite square-well potential. It has been shown that the work can be extracted through the controls of the quantum states of the particle and the width of the well in a specific manner. The "efficiency" of the engine has been calculated to be

$$\eta = 1 - \frac{E_L}{E_H}, \qquad (1)$$

where $E_H$ ($E_L$) is the value of the system energy fixed along the analog of the isothermal process at high (low) "temperature" of the cycle. (Since this engine operates without finite-temperature heat baths, it should not be confused with quantum heat engines discussed, for example, in Refs. [2-7].) It is of physical interest to examine how universal the efficiency of this form is.

To understand the physics behind the mechanism of such a Carnot-like engine, it is useful to note a structural similarity between quantum mechanics and thermodynamics. Let $H$ and $|\psi\rangle$ be the Hamiltonian and quantum state of the system under consideration, respectively. An analog of the internal energy in thermodynamics is the expectation value of the Hamiltonian,

$$E = \langle \psi | H | \psi \rangle. \tag{2}$$

This is in fact the ordinary thermodynamic internal energy in the vanishing-temperature limit, if $|\psi\rangle$ is chosen to be the ground state. Under changes of both the Hamiltonian and the state along a certain "process", it varies as $\delta E = \left( \delta \langle \psi | \right) H | \psi \rangle + \langle \psi | \delta H | \psi \rangle + \langle \psi | H \left( \delta | \psi \rangle \right)$. This has a formal analogy with the first law of thermodynamics [8]:

$$\delta' Q = \delta E + \delta' W, \tag{3}$$

where $\left( \delta \langle \psi | \right) H | \psi \rangle + \langle \psi | H \left( \delta | \psi \rangle \right)$ and $\langle \psi | \delta H | \psi \rangle$ are identified with the analogs of the changes of the quantity of heat, $\delta' Q$, and the work, $-\delta' W$, respectively. $H$ depends on the system volume, $V$, which changes in time very slowly in an equilibrium-thermodynamics-like situation. More precisely, the time scale of the change of $V$ is much larger than that of the dynamical one, $\sim \hbar/E$. Then, in the adiabatic approximation [9], holds the instantaneous Schrödinger equation, $H(V)|u_n(V)\rangle = E_n(V)|u_n(V)\rangle$, provided that the energy eigenvalues naturally satisfy the inequality $E_n(V_1) > E_n(V_2)$ for $V_1 < V_2$. Assuming that $\{|u_n(V)\rangle\}_n$ forms a complete orthonormal system, an arbitrary state $|\psi\rangle$ is expanded as $|\psi\rangle = \sum_n c_n(V)|u_n(V)\rangle$, where the expansion coefficients satisfy the normalization condition, $\sum_n |c_n(V)|^2 = 1$. Accordingly, the adiabatic scheme allows us to write

$$\delta'Q = \sum_n E_n(V)\, \delta |c_n(V)|^2, \tag{4}$$

$$\delta'W = -\sum_n |c_n(V)|^2\, \delta E_n(V). \tag{5}$$

It has been shown in Ref. [8] that imposition of the Clausius equality on the Shannon entropy (not the von Neumann entropy) and the quantity of heat makes pure-state quantum mechanics transmute into equilibrium thermodynamics at finite temperature.

Here, the following question is posed. An analog of the working material in thermodynamics is the shape of the potential that confines a particle. Then, is the efficiency in Eq. (1) universal independently of the potential?

In this paper, we answer this question by generalizing the quantum-mechanical Carnot-like engine with the one-dimensional infinite square-well potential to the case of an arbitrary confining potential. We present the most general formula for the efficiency. In marked contrast to the genuine thermodynamic Carnot engine, the efficiency of the quantum-mechanical Carnot engine depends generically on the shape of a potential as the analog of the working material, implying that the engine is not universal. This nonuniversality is due to the absence of an analog of the second law of thermodynamics in pure-state quantum mechanics, where the von Neumann entropy identically vanishes. We also identify a class of spectra that yields the efficiency of the form in Eq. (1).

Consider a cycle depicted in Fig. 1, which consists of four processes. The processes $A \to B$ and $C \to D$ are respectively analogs of the isothermal processes at high and low temperatures in classical thermodynamics, whereas $B \to C$ and $D \to A$ are those of the adiabatic processes. Each volume change is performed through control of the potential shape. An example is an optical trap of ions, in which the shape of a trapping potential can be controlled electromagnetically. Because of the absence of heat baths, quantum coherence is always kept intact. The quantum-mechanical Carnot engine is constructed employing two states, say $|u_i\rangle$ and $|u_j\rangle$. In analogy with classical thermodynamics, the "internal" energy, $E$, in Eq. (2) is kept constant during the processes $A \rightarrow B$ and $C \rightarrow D$. On the other hand, along the "adiabatic" processes $B \rightarrow C$ and $D \rightarrow A$, the states remain unchanged, as can be seen in Eq. (4). The values of the volume satisfy the following inequalities: $V_A < V_B < V_C$ and $V_C > V_D > V_A$. The cycle is reversible since pure-state quantum dynamics is reversible. Let us analyze each process in detail.

(I) Firstly, during the expansion process $A \rightarrow B$, the state changes from $|u_i(V_A)\rangle$ to $|u_j(V_B)\rangle$. In between, the system is in a superposed state, $a_1(V)|u_i(V)\rangle + a_2(V)|u_j(V)\rangle$. As mentioned above, in analogy with the isothermal expansion in classical thermodynamics, the value of the "internal" energy is fixed, i.e., $\delta E = 0$:

$$E_i(V)|a_1(V)|^2 + E_j(V)|a_2(V)|^2 \equiv E_H, \tag{6}$$

where $E_H$ is a constant independent of $V$, and $E_i(V) < E_j(V)$. From the normalization condition, it follows that

$$|a_1(V)|^2 = \frac{E_j(V) - E_H}{\Delta E(V)}, \qquad |a_2(V)|^2 = \frac{E_H - E_i(V)}{\Delta E(V)}, \tag{7}$$

where

$$\Delta E(V) \equiv E_j(V) - E_i(V). \tag{8}$$

On the other hand, the boundary condition, $a_1(V_A) = a_2(V_B) = 1$, leads to

$$E_i(V_A) = E_j(V_B) = E_H. \tag{9}$$

Since $\delta E(V) = 0$, we have $\delta'W = \delta'Q = -\Delta E(V)\delta|a_1(V)|^2 = \Delta E(V)\delta|a_2(V)|^2$.

Using Eq. (7), we find that the pressure, $P = d'W/dV$, is given as follows:

$$P_{AB}(V) = \Delta E(V) \frac{\partial}{\partial V}\left[\frac{E_H - \bar{E}(V)}{\Delta E(V)}\right], \tag{10}$$

where

$$\bar{E}(V) \equiv \frac{E_i(V) + E_j(V)}{2}. \tag{11}$$

Therefore, the work in the process is

$$Q_H \equiv W_{AB} = \int_{V_A}^{V_B} dV\, \Delta E(V) \frac{\partial}{\partial V}\left[\frac{E_H - \bar{E}(V)}{\Delta E(V)}\right]. \tag{12}$$

Note that Eq. (9) enables us to express this quantity in terms only of $V_A$.

(II) Next, during the expansion process $B \rightarrow C$, $\delta'Q = 0$, and the system remains in the state, $|u_j\rangle$. So, the pressure reads $P_{BC}(V) = -\partial E_j(V)/\partial V$. It is necessary to assume that $P_{BC}(V)$ decreases faster than $P_{AB}(V)$ in Eq. (10) as $V$ increases. However,

this point may always be fulfilled since the energy eigenvalues decrease with respect to $V$. The work is calculated to be $W_{BC} = E_j(V_B) - E_j(V_C)$. From Eq. (9), it is rewritten as

$$W_{BC} = E_i(V_A) - E_j(V_C). \tag{13}$$

(III) Then, during the compression process $C \to D$, the state changes from $|u_j(V_C)\rangle$ to $|u_i(V_D)\rangle$. In between, the system is in a superposed state, $b_1(V)|u_i(V)\rangle + b_2(V)|u_j(V)\rangle$, as in (I). The value of the "internal" energy is fixed:

$$E_i(V)|b_1(V)|^2 + E_j(V)|b_2(V)|^2 \equiv E_L, \tag{14}$$

where $E_L$ is a constant independent of $V$. From the normalization condition, it follows that

$$|b_1(V)|^2 = \frac{E_j(V) - E_L}{\Delta E(V)}, \qquad |b_2(V)|^2 = \frac{E_L - E_i(V)}{\Delta E(V)}. \tag{15}$$

On the other hand, the boundary condition, $b_2(V_C) = b_1(V_D) = 1$, leads to

$$E_j(V_C) = E_i(V_D) = E_L. \tag{16}$$

Similarly to (I), the pressure is found to be

$$P_{CD}(V) = \Delta E(V) \frac{\partial}{\partial V}\left[\frac{E_L - \bar{E}(V)}{\Delta E(V)}\right]. \tag{17}$$

Therefore, the work is given by

$$W_{CD} = \int_{V_C}^{V_D} dV\, \Delta E(V) \frac{\partial}{\partial V}\left[\frac{E_L - \bar{E}(V)}{\Delta E(V)}\right]. \tag{18}$$

Eq. (16) allows us to express this quantity in terms only of $V_C$.

(IV) Lastly, during the compression process $D \rightarrow A$, the system remains in the state, $|u_i\rangle$. The pressure reads $P_{DA}(V) = -\partial E_i(V)/\partial V$. Accordingly, the work is given by $W_{DA} = E_i(V_D) - E_i(V_A)$. From Eq. (16), it is rewritten as

$$W_{DA} = E_j(V_C) - E_i(V_A). \tag{19}$$

Note that the relation $W_{DA} = -W_{BC}$ holds, as in the case of the genuine Carnot cycle in classical thermodynamics. Therefore, the total work extracted after one cycle is $W = W_{AB} + W_{BC} + W_{CD} + W_{DA} = W_{AB} + W_{DA}$. Consequently, we obtain the following most general expression for the efficiency, $\eta = W/Q_H$:

$$\eta = 1 - \frac{\int_{V_D}^{V_C} dV\, \Delta E(V) \frac{\partial}{\partial V}\left[\frac{E_L - \bar{E}(V)}{\Delta E(V)}\right]}{\int_{V_A}^{V_B} dV\, \Delta E(V) \frac{\partial}{\partial V}\left[\frac{E_H - \bar{E}(V)}{\Delta E(V)}\right]}, \tag{20}$$

which can be expressed only in terms of the smallest and largest values of the volume, $V_A$ and $V_C$, because of Eqs. (9) and (16). The formula in Eq. (20) is the main result of the present work. It holds not only for single-particle systems but also for many-particle systems. Note that $\eta < 1$, since the second term on the right-hand side is not zero, in general. However, a condition for the maximum value of $\eta$ is yet to be discovered.

Among the four processes, two of them, (I) and (III), are highly nontrivial, in practice. It is necessary to precisely realize those superposed states during the expansion and compression processes, which may require fine quantum-state engineering.

Below, we evaluate the efficiency in Eq. (20) for several examples.

Firstly, let us reexamine the infinite square-well potential in one dimension discussed in Ref. [1]. The energy eigenvalues are given by $E_n(L) = n^2 \pi^2 \hbar^2 / (2mL^2)$ ($n = 1, 2, 3, ...$), where $m$ and $L$ are the mass of the particle confined in the potential and the slowly varying width of the well, respectively. The volume, $V$, corresponds to the width, $L$, in one dimension. Accordingly, the pressure corresponds to the force. Take the energies of the ground and first excited states, $E_1(L)$ and $E_2(L)$. From Eqs. (9) and (16), where $V$ should be replaced by $L$, we have that $L_B = 2L_A$, $L_D = L_C / 2$, $E_H = \pi^2 \hbar^2 / (2mL_A^2)$, and $E_L = 4\pi^2 \hbar^2 / (2mL_C^2)$. Eq. (20) can immediately be evaluated for this system to yield Eq. (1). In terms of the values of the width, the efficiency is given by $\eta = 1 - 4(L_A / L_C)^2$. These are precisely the results for the quantum-mechanical Carnot-like engine obtained earlier in Ref. [1]. We would like to mention the following recent studies. In Ref. [10], a discussion about a finite-time process has been developed, and the value of this efficiency is rigorously determined in the case when the power output is maximum. In Ref. [11], it has been shown how superposition of quantum states can enhance the efficiency of this engine.

Next, let us consider a particle with mass $m$ confined in the one-dimensional harmonic oscillator potential, $U(x) = (1/2)kx^2$, where $k$ is a positive factor that can vary slowly. At the value, $U(x) = U_0$, the width of the potential is $L = \sqrt{8U_0 / k}$, and thus the frequency depends on $L$ as $\omega(L) = \sqrt{k/m} = \sqrt{8U_0/m}/L$. The energy eigenvalues are $E_n(L) = \hbar\omega(L)(n+1/2)$ ($n = 0, 1, 2, ...$). Take the energies of the ground and first excited states, $E_0(L)$ and $E_1(L)$. It follows from Eqs. (9) and (16) that $L_B = 3L_A$, $L_D = L_C/3$, $E_H = (\hbar/2)\sqrt{8U_0/m}/L_A$, and $E_L = (3\hbar/2)\sqrt{8U_0/m}/L_C$. Therefore, again, Eq. (20) is reduced to Eq. (1). In terms of the values of the width, the efficiency is given by $\eta = 1 - 3L_A/L_C$.

Now, let us examine a class of energy spectra. Suppose that the energy eigenvalues have the form [12]

$$E_n(V) = \frac{\varepsilon_n}{V^\alpha} \qquad (n = 1, 2, 3, ...), \qquad (21)$$

where $\alpha$ and $\varepsilon_n$'s are independent of $V$, and, in particular, $\alpha$ is positive and $\varepsilon_n$'s satisfy $\varepsilon_1 < \varepsilon_2 < \varepsilon_3 < \cdots$. A class of spectra of this form is referred to here as *homogeneous type*. Take the first two states, for example. Then, Eqs. (9) and (16) lead to $V_B = (\varepsilon_2/\varepsilon_1)^{1/\alpha} V_A$, $V_D = (\varepsilon_1/\varepsilon_2)^{1/\alpha} V_C$, $E_H = \varepsilon_1/V_A^\alpha$, and $E_L = \varepsilon_2/V_C^\alpha$. Consequently, Eq. (20) is calculated to result in Eq. (1). In other words, the efficiency of the form in Eq. (1) is "universal" within the class of homogeneous-type spectra. In terms of the values of the volume, the efficiency is given by $\eta = 1 - (\varepsilon_2/\varepsilon_1)(V_A/V_C)^\alpha$. In view of this, the above-mentioned examples of the infinite square well and harmonic oscillator, both of which are of the homogeneous type, simply correspond to $\alpha = 2$ and $\alpha = 1$ (with $V$ being replaced by $L$), respectively.

Finally, we discuss the Morse potential in three dimensions. It reads

$U(r) = D\{\exp[-2\alpha(r-r_0)] - 2\exp[-\alpha(r-r_0)]\}$, where $D$, $\alpha$, and $r_0$ are positive factors, and $r$ is the radial part of the spherical coordinate. This potential has the minimum, $-D$, at $r = r_0$. The potential width, $R$, at the value $U = -U_0$ ( $0 < U_0 < D$ ) is $R = (1/\alpha) \ln(\rho_+/\rho_-)$, where $\rho_\pm \equiv 1 \pm \sqrt{1 - U_0/D}$. In the case of the vanishing angular-momentum states (i.e., the *s* states), the energy eigenvalues of the bound states are given as follows [13]: $E_n = -D + \hbar\alpha\sqrt{2D/m}\,(n+1/2) - [\hbar^2\alpha^2/(2m)](n+1/2)^2$, where $n = 0, 1, 2, ... < \sqrt{2mD}/(\hbar\alpha) - 1/2$, which is not a homogeneous type in Eq. (21) with $V$ being replaced by $R$. Let us take the energies of the ground and first excited states: $E_0(R) = -D + \kappa/R - \lambda/R^2$ and $E_1(R) = -D + 3\kappa/R - 9\lambda/R^2$, where $\kappa = \hbar\sqrt{D/(2m)}\,\ln(\rho_+/\rho_-)$ and $\lambda = [\hbar^2/(8m)][\ln(\rho_+/\rho_-)]^2$, respectively. $R$ varies as in figure 1 (with $V$ being replaced by $R$). The system is "breathing". Eqs. (9) and (16) lead to $R_B = 3R_A$ and $R_D = R_C/3$. The efficiency is found to be $\eta = 1 - F/G$, where

$F = (E_L + D)\ln 3 - [E_L + D - 3\kappa^2/(16\lambda)]\ln|(R_C - 4\lambda/\kappa)/(R_C - 12\lambda/\kappa)| - 3\kappa/(2R_C)$ and

$G = (E_H + D)\ln 3 - [E_H + D - 3\kappa^2/(16\lambda)]\ln|[R_A - 4\lambda/(3\kappa)]/(R_A - 4\lambda/\kappa)| - \kappa/(2R_A)$.

Clearly, this does not have the form in Eq. (1).

In conclusion, we have derived the most general formula for the efficiency of the pure-state quantum-mechanical Carnot-like engine reversibly operating at vanishing temperatures. We have shown how the efficiency depends on the shape of a potential as the analog of the working material in thermodynamics. Thus, this engine is generically not universal, in contrast to the genuine thermodynamic Carnot engine. This nonuniversality is due to the absence of an analog of the second law of thermodynamics in pure-state quantum mechanics, in which the von Neumann entropy identically vanishes. We have also identified a class of spectra of the homogeneous type, which has the efficiency of the common form in Eq. (1).

**Acknowledgment**

This work was supported in part by a Grant-in-Aid for Scientific Research from the Japan Society for the Promotion of Science.

**References**

1  Bender, C.M.; Brody, D.C.; Meister, B.K. Quantum mechanical Carnot engine. *J. Phys. A: Math. Gen.* **2000**, *33*, 4427-4436.

2. Geusic, J.E.; Schulz-DuBois, E.O.; and Scovil, H.E.D. Quantum equivalent of the Carnot cycle. *Phys. Rev.* **1967**, *156*, 343-351.

3. Kieu, T.D. The second law, Maxwell's demon, and work derivable from quantum heat engines. *Phys. Rev. Lett.* **2004**, *93*, 140403–1-4.

4. Quan, H.T.; Zhang, P.; Sun, C.P. Quantum heat engine with multilevel quantum systems. 2005 *Phys. Rev. E* **2005**, *72*, 056110–1-10.

5. Johal, R.S. Quantum heat engines and nonequilibrium temperature. *Phys. Rev. E* **2009**, *80*, 041119–1-5.

6. Beretta, G.P. Quantum thermodynamic Carnot and Otto-like cycles for a two-level system. *EPL* **2012**, *99*, 20005–1-5.

7. Nakamura, K.; Sobirov, Z.A.; Matrasulov, D.U.; Avazbaev, S.K. Bernoulli's formula and Poisson's equations for a confined quantum gas: Effects due to a moving piston. *Phys. Rev. E* **2012**, *86*, 061128–1-10.

8. Abe, S.; Okuyama, S. Similarity between quantum mechanics and thermodynamics: Entropy, temperature, and Carnot cycle. *Phys. Rev. E* **2011**, *83*, 021121–1-3.

9. Migdal, A.B. *Quantitative Methods in Quantum Theory*; Addison-Wesley: New York, NY, USA, 1989.

10. Abe, S. Maximum-power quantum-mechanical Carnot engine. *Phys. Rev. E* **2011**, *83*, 041117–1-3.

11. Abe, S.; Okuyama, S. Role of the superposition principle for enhancing the efficiency of the quantum-mechanical Carnot engine. *Phys. Rev. E* **2012**, *85*, 011104–1-4.

12. Wang, J.; He, J.; He, X. Performance analysis of a two-state quantum heat engine working with a single-mode radiation field in a cavity. *Phys. Rev. E* **2011**, *84*, 041127–1-6.

13. Flügge, S. *Practical Quantum Mechanics*; Springer-Verlag: Berlin, Germany, 1994.

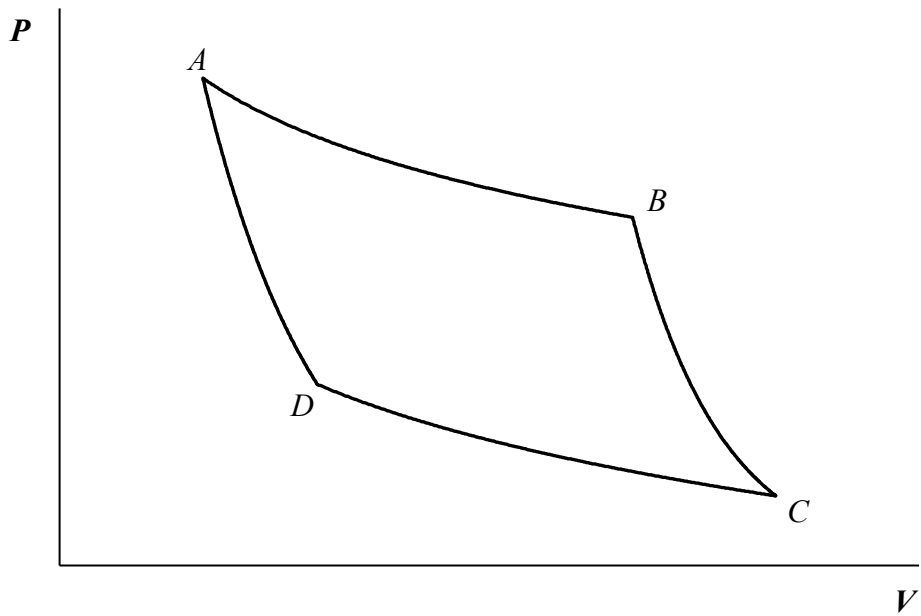

**Figure 1.** The quantum-mechanical analog of the Carnot cycle depicted in the plane of the volume $V$ and pressure $P$. During $A \to B$ and $C \to D$, the expectation values of the Hamiltonian are fixed, whereas during $B \to C$ and $D \to A$, the quantum states are kept unchanged.